# Collective Excitations of $^{154}$Sm nucleus at FEL γ ⊗ LHC Collider

E. GULIYEV[1], A. A. KULIEV[2,3], S. SULTANSOY[3,4], Ö. YAVAŞ[1]

[1] *Department of Eng. Physics, Ankara University, 06100 Tandogan, Ankara, Turkey*

[2] *Department of Physics, Faculty of Arts and Sciences, Sakarya University, 54100 Serdivan, Adapazarı, Turkey*

[3] *Institute of Physics, Academy of Sciences, H. Cavid avenue 33, Baku, Azerbaijan*

[4] *Department of Physics, Faculty of Arts and Sciences, Gazi University, 06500 Ankara Turkey*

**Abstract.** – The production of collective excitations of the $^{154}$Sm at FEL γ ⊗ LHC collider is investigated. We show that this machine will be a powerful tool for investigation of high energy level excitations.

During the last decade great success has been achieved in measurement of the nuclear excitations with low multipolarity [1]. The study of these excitations, in particular collective magnetic dipole ones, gives valuable information about nuclear structure and nucleon-nucleon forces at low energy. The interest in studying these excitations has increased concerning the possibility of pion precondensations and Δ isobar contributions to M1 response [2] and, besides, it has important implication on astrophysical processes [3]. A low lying branch of these excitations (ω < 4 MeV), so called scissors mode states, which are predominantly excited by orbital part of magnetic dipole operator, has been found for the isotopes with permanent deformation in a wide region beginning from light nuclei up to the actinides, including also transitional and γ-soft nuclei [4]. While spin-isospin interactions are responsible for generation of spin-vibrational magnetic dipole resonance in the region of the neutron binding energy [5], experimentally the identification of high energy branch of the spin response is far from being trivial: measurements are made difficult by the weakness of the resonance and fragmentation of its strength in the low energy tail of the giant dipole resonance below neutron emission threshold (E < 9 MeV).

A comprehensive study of the magnetic dipole response in (e, e') scattering at upward angle [6], (p, p') scattering at extreme forwarded angle [7,8] and using (γ, γ') technique in tagget photon experiment [9] show that in heavy nuclei there exists very broad M1 resonance at energy interval between 7 and 11 MeV (centred at E ≈ 44·A$^{-1/3}$ MeV). Moreover, there is the major disagreement between electron and proton scattering experiments, the broad structure observed in (p, p') measurement has not been seen in the electron scattering. Indeed, because of the limited energy resolution, individual states could not be discerned in the existing scattering experiments. Opportunities provided by the existing tagged photon sources also are not sufficient for studying these levels because energy resolution is ~50 keV in multi-MeV region [10], therefore it is not sufficient for studying individual levels.

After the discovery of scissors mode excitations [11], they were registered in many deformed nuclei using Nuclear Resonance Fluorescence (NRF) method [12]. However, capacity of NRF experiments for study of higher-lying excitations are limited by many factors [13]:

- Neutron induced γ background from electron set-up material collisions: (e, n) and, subsequently, (n, nγ). The dramatically increased continuous background and the discrete lines from (n, γ) reactions do not allow a safe interpretation of the (γ, γ') spectrum anymore.

- Rather weak interaction between the photons and the target material. For typical obtainable photon currents of about $10^6$ γ/(s keV) one needs around 500 mg isotopically enriched target material.

- The typical count rates in the Ge(HP) detectors is an order of a few thousand events per second at electron beam currents of about 5 µA leading to typical measuring times of several hours for the spectra 4-8 MeV.

Recently, a new method has been proposed to investigate the scattering reactions with photons on fully ionised nuclei, namely FEL γ-nucleus colliders [14,15], which is free from above-mentioned troubles. To induce γ-nucleus collisions a Free Electron Laser (FEL) and a heavy ion synchrotron have been considered bringing them together (see Fig.1 in Ref. [15]).

This facility allows to produce monochromatic photons with high density and the accelerated fully ionized nuclei see the few keV energy FEL photons as a "laser" beam with MeV energy. Due to good monohromaticity ($\Delta E/E < 10^{-3} \div 10^{-4}$) with typical obtainable number of photons in order $10^{13}$γ/bunch and excellent tunability, this method can be successfully used to investigate nuclear excitations with low multipolarity in a wide energy region. The potential of FEL γ-nucleus colliders in search for collective $I^\pi=1^+$ excitations of $^{140}$Ce nucleus was investigated in [16]. In this paper we study the production of high-energy excitations of $^{154}$Sm nucleus at FEL γ ⊗ LHC collider.

In Table I we present main characteristics of $^{154}$Sm $1^+$ excitations observed in (γ,γ') scattering [17]. It is seen that observed levels lie between 2 and 4 MeV. Of course, a lot of higher energy levels should exist, but they can not be detected by present experiments for reasons given above. In Table 2 characteristics of $^{154}$Sm excitations in 4-10 MeV region are presented in the framework of the rotational RPA model of Ref [18]. In order to demonstrate the advantage of the new method, we choose this region, because it can not be explored in details by exiting and foreseeable methods.

Concerning FEL γ ⊗ LHC collider, the energy of FEL photons needed for excitation of corresponding $^{154}$Sm level can be expressed as (for details, see [15, 16]):

$$\omega_0 = \frac{E_{exc}}{2\gamma_{Sm}} = \frac{A \cdot E_{exc}}{2Z \cdot \gamma_p}$$

where $E_{exc}$ is the energy of corresponding excited level, $Z$ and $A$ are atomic and mass numbers of nucleus, $\gamma_p$ is the Lorentz factor of the proton. Taking into account, that, Z=62 and A=154 for the considered isotope of Sm and $\gamma_p = 7TeV/m_p \approx 7462$ for LHC, we obtain $\omega_0 \approx 1.6642 \cdot 10^{-4} \cdot E_{exc}$. Corresponding values for needed FEL photon energies are given in the second column of Tables I and II. We see that energies of FEL photons lie in the region 0.43 keV÷1.59 keV, therefore, reasonable upgrade of the TTF FEL [19] parameters (see Table IV) cover needed region.

Indeed, the energy of FEL photons can be expressed in practical units as [20]

$$\omega[eV] = 950 \frac{(E_e[GeV])^2}{\lambda_u[cm] \cdot (1 + K^2/2)}$$

where $\lambda_u$ is the period length of the undulator, $E_e$ is the energy of the electron beam, $K = 0.934 \cdot B_0[T] \cdot \lambda_u[cm]$ is the strength parameter of the undulator and $B_0$ is the peak value of magnetic field. It is clear that $E_{FEL}$ may be adjusted by changing $E_e$, $B_0$ or $\lambda_u$.

Luminosity of the FEL γ-Sm nucleus collisions can be expressed as

$$L = \frac{n_\gamma n_{Sm}}{4\pi\sigma_x\sigma_y} f_{coll}$$

where $n_\gamma$ and $n_{Sm}$ are the number of particles in photon and nucleus bunches, respectively, $\sigma_x$ and $\sigma_y$ are the transverse beam sizes and $f_{coll}$ is the collision frequency. Using the values from Tables III and IV, luminosity is obtained as L=2·10$^{30}$ cm$^{-2}$s$^{-1}$ for FEL γ-$^{154}$Sm collision at LHC.

The cross section for the resonant photon scattering is given by the Breit-Wigner formula

$$\sigma_{res}(\gamma,\gamma') = \frac{\pi}{E^2} \frac{2J_{exc}+1}{(2J_0+1)} \frac{B_{in}B_{out}\Gamma^2}{(E-E_R)^2+\Gamma^2/4}$$

where E is the c.m. energy of the incoming photon (in our case it is very close to that in the rest frame of the nucleus), $J_{exc}$ and $J_0$ are spins of the excited and ground states of the nucleus. Spin of the ground states of $^{154}$Sm is equal to zero and $J_{exc}=1$ for excitations under consideration. $B_{in}$ and $B_{out}$ are branching fractions of the excited nucleus into the entrance and exit channels, respectively, $E_R$ is the energy at the resonance and $\Gamma$ is the total width of the excited states. The approximate value of averaged cross section has been found to be $\sigma_{ave} \cong \sigma_{res}\Gamma/\Delta E_\gamma$, where $\Delta E_\gamma$ is the energy spread of FEL beam in the nucleus rest frame; $\Delta E_\gamma/E_\gamma \leq 10^{-3}$, $E_\gamma \cong E_{exc}$ [15]. Event numbers for the excited levels can be calculated from $R=L \cdot \sigma_{ave}$ using given luminosity and average cross section for LHC. Calculated event numbers are given in the last column of Tables I and II.

Finally, the number of $(\gamma, \gamma')$ events are of the order of $10^7$ per second for considered excitations of $^{154}$Sm isotope. Using this huge statistics one will be able to investigate a lot of isotopes in a short time interval.

Table I Integral characteristics and (γ,γ') scattering results for $1^+$ excitations of $^{154}$Sm

| $E_{exc}$, MeV | $\omega_{FEL}$, keV | Γ, eV | $\sigma_{res}$, cm$^2$ | $\sigma_{ave}$, cm$^2$ | R/s, $10^7$ |
|---|---|---|---|---|---|
| 2.555 | 0.425 | 0.030 | 0.112·10$^{-20}$ | 0.133·10$^{-22}$ | 2.66 |
| 2.616 | 0.435 | 0.036 | 0.107·10$^{-20}$ | 0.146·10$^{-22}$ | 2.92 |
| 3.091 | 0.514 | 0.052 | 0.768·10$^{-21}$ | 0.128·10$^{-22}$ | 2.56 |
| 3.193 | 0.531 | 0.101 | 0.720·10$^{-21}$ | 0.228·10$^{-22}$ | 4.56 |
| 3.826 | 0.636 | 0.048 | 0.501·10$^{-21}$ | 0.634·10$^{-23}$ | 1.27 |

Table II Integral characteristics and theoretical results for $1^+$ excitations of $^{154}$Sm

| $E_{exc}$, MeV | $\omega_{FEL}$ keV | Γ, eV | $\sigma_{res}$, cm$^2$ | $\sigma_{ave}$, cm$^2$ | R/s, $10^7$ |
|---|---|---|---|---|---|
| 4,600 | 0.765 | 0,077 | 0,347 10$^{-21}$ | 0,585 ·10$^{-23}$ | 1.17 |
| 4,728 | 0.786 | 0,081 | 0,328 10$^{-21}$ | 0,563 ·10$^{-23}$ | 1.13 |
| 5,806 | 0.966 | 0,173 | 0,218 10$^{-21}$ | 0,649·10$^{-23}$ | 1.30 |
| 6,278 | 1.044 | 0,290 | 0,186 10$^{-21}$ | 0,861 ·10$^{-23}$ | 1.72 |
| 6,797 | 1.131 | 1,170 | 0,159 10$^{-21}$ | 0,274 ·10$^{-22}$ | 5.48 |
| 7,237 | 1.204 | 0,789 | 0,140 10$^{-21}$ | 0,153 ·10$^{-22}$ | 2.60 |
| 7,443 | 1.238 | 0,695 | 0,132 10$^{-21}$ | 0,124 ·10$^{-22}$ | 2.48 |
| 8,465 | 1.408 | 1,940 | 0,102 10$^{-21}$ | 0,234 ·10$^{-22}$ | 4.68 |
| 9,049 | 1.505 | 2,180 | 0,898 10$^{-22}$ | 0,217 ·10$^{-22}$ | 4.34 |
| 9,138 | 1.520 | 5,700 | 0,879 10$^{-22}$ | 0,549 ·10$^{-23}$ | 1.10 |
| 9,438 | 1.587 | 5,670 | 0,824 10$^{-22}$ | 0,495 ·10$^{-22}$ | 9.90 |
| 9,451 | 1.572 | 4,610 | 0,822 10$^{-22}$ | 0,401 ·10$^{-22}$ | 8.02 |
| 9,583 | 1.594 | 2,880 | 0,799 10$^{-22}$ | 0,240 ·10$^{-22}$ | 4.80 |

Table III. Main parameters of the TTF FEL beam

| | |
|---|---|
| Number of electrons per bunch $n_e$, $10^{10}$ | 1 |
| Pulse length, μs | 1000 |
| Number of bunches per pulse $n_b$ | 7200 |
| Repetition rate, Hz | 10 |
| Number of photons per bunch $n_\gamma$, $10^{13}$ | 4 |
| Energy of FEL photons ω, eV | 193 |
| Energy spread of photons $\Delta E_{FEL}/E_{FEL}$, $10^{-3}$ | 1 |
| RMS beam size $\sigma_{x,y}$, μm | 50 |

Table IV. Main parameters of $^{154}$Sm beam at LHC

| | $^{154}$Sm at LHC |
|---|---|
| Maximum beam energy $E_{Sm}$, TeV | 434 |
| Particles per bunch $n_{nuc}$, $10^8$ | 1.4 |
| Normalized emittance $\varepsilon^N$, mm mrad | 1.5 |
| Amplitude function at IP $\beta^*$, cm | 20 |
| RMS beam size at IP $\sigma_{x,y}$, μm | 10 |
| Bunch spacing, ns | 100 |
| Number of bunches in FEL pulse $n_b$ | 10000 |